\documentclass[10pt]{article}

\usepackage{amssymb,amsfonts,amsmath,bm}
\usepackage{graphicx}
\usepackage[usenames]{color}

\usepackage{xr}

\usepackage{array}

\usepackage{color}

\usepackage{cite}

\usepackage{capt-of}
\usepackage{rotating}

\catcode`\_=\active
\def_#1{\ensuremath{\sb{\text{#1}}}}

\usepackage{setspace} 
\usepackage[labelfont=bf,labelsep=period,justification=raggedright]{caption}

\makeatletter
\renewcommand{\@biblabel}[1]{\quad#1.}
\makeatother

\date{}

\begin{document}

\begin{flushleft}
{\Large
\textbf{Thermodynamics of sustaining gases in the roughness of submerged superhydrophobic surfaces}
}
\\
Neelesh A. Patankar$^{\ast}$\\

Department of Mechanical Engineering, Northwestern University, 
\\
2145 Sheridan Road, Evanston, IL 60208, USA

$^\ast$ E-mail: n-patankar@northwestern.edu
\end{flushleft}


\section*{Abstract}
Rough surfaces submerged in a liquid can remain almost dry if the liquid does not fully wet the roughness and gases are sustained in roughness grooves. Such partially dry surfaces can help reduce drag or enhance boiling. Gases sustained in roughness grooves would be composed of air and the vapor phase of the liquid itself. The thermodynamics of sustaining vapor was considered in a prior work  [Patankar, \textit{Soft Matter}, 2010, \textbf{6}, 1613]. Here, the thermodynamics of sustaining gases (e.g. air) is considered. Governing equations are presented along with a solution methodology to determine a critical condition to sustain gases. The critical roughness scale to sustain gases is estimated for different degrees of saturation of gases dissolved in the liquid. It is shown that roughness spacings of less than a micron are essential to sustain gases on surfaces submerged in water at atmospheric pressure. This is consistent with prior empirical data.  

\section{Introduction}
Rough surfaces that exhibit non-wetting properties under submerged conditions are desirable for many applications like drag reduction,\cite{Gova09a, Leec11a, Mcha09a} boiling,\cite{Dhir06a} among others. Here, non-wetting behavior is defined as the one where the liquid (e.g. water), into which the rough surface is submerged, does not fully wet the surface. For this to happen, it is essential to sustain gases in roughness grooves of the surface. However, sustaining gases in roughness grooves over long time periods has been challenging. Typically, gases are found to deplete after 2-3 days from rough surfaces with tens of micron scale features,\cite{Gova09a, Leec11a, Poet10a, Sama11a} whereas submicron scale roughness is found to sustain gases for more than 120 days.\cite{Poet10a, Balm11a} Increased pressure is found to adversely affect the ability to sustain gas.\cite{Poet10a} An understanding of the governing equations is essential to design surfaces that can sustain gases over an extended period.\cite{Marm06a, Flyn08a, Vaka13a}     

In general, gases sustained in roughness grooves would be a mixture of air and the vapor phase of the submerging liquid. Thermodynamics for sustaining vapor has been studied before. \cite{Pata10a, Pata10b, Vaka12a, Jone14a} In this work, thermodynamics of sustaining air is considered. Specifically, the goal is to present an analysis of the equilibrium state where a gas like air prefers to remain in roughness grooves under submerged conditions. In this analysis it is assumed that vapor produced by the submerging liquid is not present in roughness grooves. However, generalization is possible by using the concept of partial pressure.\cite{Mode83a} The presence of the vapor would make the surface more non-wetting.     

In the next section, background on the thermodynamics of gas dissolution is presented.  Following that, the theory is applied to an example problem of sustaining air in cylindrical pores. 

\section{Thermodynamics of gas dissolution}

\subsection{Henry's law}

\begin{figure}[h]
\centering
\includegraphics[height=5cm]{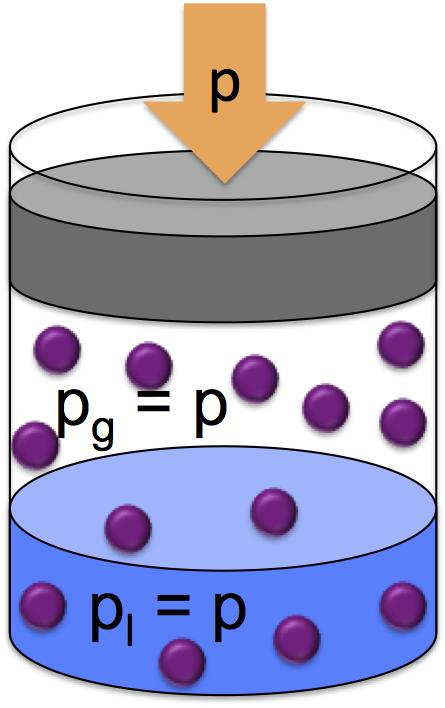}
\caption{Chemical equilibrium between a gas (denoted by circles) dissolved in a liquid solvent and the same gas outside the solvent. The liquid solvent is assumed to be non-volatile.}
\label{fig-henry-law}
\end{figure}

Consider a gas in chemical equilibrium with a liquid bath into which this same gas is dissolved (Fig.~\ref{fig-henry-law}). Consider the liquid solvent to be non-volatile, i.e., it does not evaporate. This implies that there will be only one type of gas present above the liquid bath. 
It is generally observed that the greater the pressure of the gas, the greater will be the dissolved mole fraction of the gas in the liquid. The equilibrium pressure (of the gas outside the liquid) versus the mole fraction (of the gas dissolved in the liquid) relationship is given by Henry's law:
\begin{equation}
	\frac{p_g}{x_{g,l}} = H^c_{g,l},
	\label{eqn-henry-law}
\end{equation}
where $p_g$ is the pressure of the gas outside the liquid, $x_{g,l}$ is the mole fraction of the gas dissolved in the liquid, and $H^c_{g,l}$ is Henry's constant. Eqn.~\ref{eqn-henry-law} is valid for ideal solutions. Real solutions have non-idealities due to which the pressure to mole fraction ratio is not constant in general. 

\subsection{Chemical equilibrium}

An expression for the pressure-mole fraction relationship at equilibrium is obtained by equating the chemical potential of the gas outside the liquid to the chemical potential of the gas dissolved in the liquid. 

The chemical potential $\mu_{g}$ of the gas outside the liquid 
is given by\cite{Mode83a, Ludw05a} 
\begin{equation}
	\mu_{g} = \mu^o_{g}[T] + RT~\textup{ln}[\psi_{g}p_g],
	\label{eqn-cp-gas}
\end{equation}
where $[\cdot]$ implies ``function-of," $\mu^o_g$ is the standard potential of the gas which depends only on temperature $T$, $R$ is the universal gas constant, $\psi_g$ is the fugacity coefficient of the gas that accounts for non-idealities, and $p_g$ is the pressure of the gas. 
In general, $\psi_g$ is not constant.

The chemical potential $\mu_{g,l}$ of the dissolved gas is given by\cite{Mode83a, Ludw05a}
\begin{equation}
	\mu_{g,l} = \mu^*_{g,l}[p_l,T] + RT~\textup{ln}[\gamma_{g,l} x_{g,l}],
	\label{eqn-cp-solute}
\end{equation}
where $\gamma_{g,l}$ is the activity coefficient, $x_{g,l}$ is the mole fraction, and $\mu^*_{g,l}[p_l,T]$ is the standard potential of the dissolved gas in the liquid that has pressure $p_l$ and temperature $T$. In general, $\gamma_{g,l}$ is not constant. 

At equilibrium the chemical potentials of the gas phase and the dissolved gas must be equal. It follows from Eqns.~\ref{eqn-cp-gas} and~\ref{eqn-cp-solute} that 
\begin{equation}
	\frac{p_{g}}{x_{g,l}} = \frac{\gamma_{g,l}}{\psi_g} e^{\frac{\mu^*_{g,l}[p_l,T] - \mu^o_{g}[T]}{RT}}.
	\label{eqn-eqlm}
\end{equation}
Eqn.~\ref{eqn-eqlm} is a general expression where the pressure $p_g$ of the gas phase and the pressure $p_l$ of the liquid need not be equal. For the case depicted in Fig.~\ref{fig-henry-law}, $p_g$ and $p_l$ are indeed equal since liquid solvent is assumed to be non-volatile. However, consider another situation where there are multiple gas species above the liquid in Fig.~\ref{fig-henry-law}. In this case, $p_g$ would be the partial pressure of the gas of interest and $p_l$ would be equal to the total pressure exerted by all gases on the liquid.

If the gas is assumed to be ideal, then $\psi_g = 1$. Additionally, if the solution is dilute, which is often the case with gases dissolved in liquids, then $\gamma_{g,l} \rightarrow 1$. Using this in Eqn.~\ref{eqn-eqlm}
\begin{equation}
	\frac{p_{g}}{x_{g,l}} = H[p_l^{1atm}, T]e^{\frac{\mu^*_{g,l}[p_l,T] - \mu^*_{g,l}[p_l^{1atm},T]}{RT}},
	\label{eqn-eqlm-rearr}
\end{equation}
where $p_l^{1atm}$ implies that $p_l = 1$~atmosphere (atm.), which is chosen to be the reference pressure. Additionally,
\begin{equation}
	H[p_l^{1atm}, T] = e^{\frac{\mu^*_{g,l}[p_l^{1atm},T] - \mu^o_{g}[T]}{RT}}.
	\label{eqn-H}
\end{equation}
It follows from Eqn.~\ref{eqn-eqlm-rearr} that $p_g/x_{g,l} = H[p_l^{1atm}, T]$ when $p_l = 1$~atm. Hence, $H[p_l^{1atm}, T]$ is Henry's constant (see Eqn.~\ref{eqn-henry-law}) at 1~atm. pressure of the liquid.
It is seen from Eqn.~\ref{eqn-eqlm-rearr} that the pressure to mole fraction ratio $p_g/x_{g,l}$ is dependent on the temperature, as well as the pressure of the liquid.\cite{Ludw05a} 
Eqn.~\ref{eqn-eqlm-rearr} can be simplified further to obtain\cite{Mode83a}
\begin{equation}
	\frac{p_{g}}{x_{g,l}} = H[p_l^{1atm}, T] e^{\frac{V_{g,l}}{RT}(p_l-p_l^{1atm})},
	\label{eqn-eqlm-ideal}
\end{equation}
where $V_{g,l}$ is the partial molar volume of the gas dissolved in the liquid. In general, $V_{g,l}$ is dependent on $x_{g,l}$, $T$, and $p_l$. However, the variation of $V_{g,l}$ is not significant even upto liquid pressures as high as 1000~atm.\cite{Ludw05a}

\subsection{Isobaric equilibrium}
\label{isobar}

\begin{figure}[h]
\centering
\includegraphics[height=6cm]{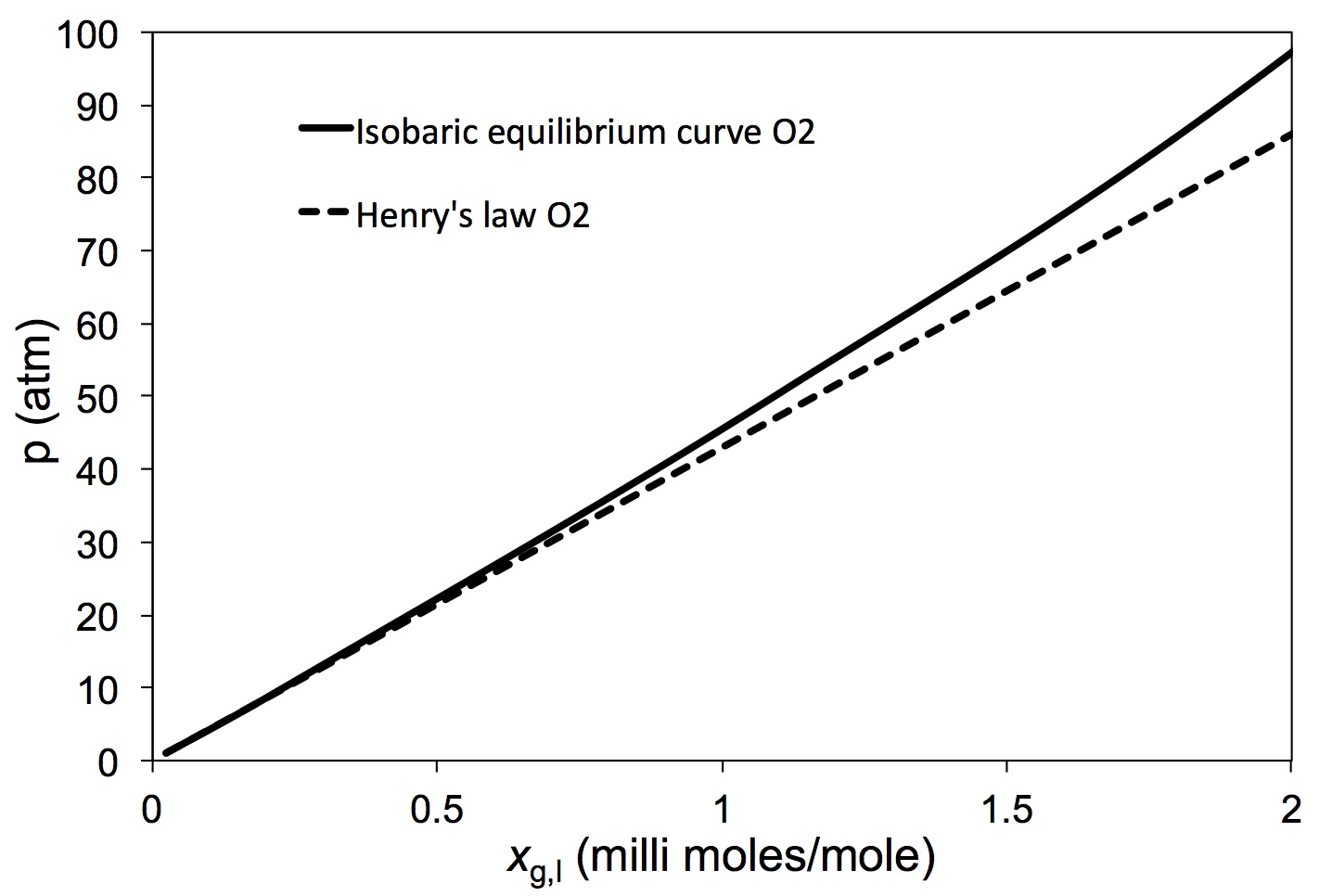}
\caption{The isobaric equilibrium curve (Eqn.~\ref{eqn-coexist}) compared to the equilibrium relation based on Henry's law (Eqn.~\ref{eqn-henry-law}). 
The gas is oxygen and the solvent is water, where $H_{g,l}^c = H[p_l^{1atm}, T = 25^\textup{o}\textup{C}]$ $=42,590$~atm, and $V_{g,l} = 32$~ml/mol.\cite{Ludw05a}}
\label{fig-eqlm-curve}
\end{figure}

For the case depicted in Fig.~\ref{fig-henry-law}, $p_{g} = p_l = p_e$, where $p_e$ is the equilibrium pressure. Since the gas and liquid pressures are equal, this will be regarded as an isobaric equilibrium. 
For isobaric equilibrium, Eqn.~\ref{eqn-eqlm-ideal} becomes 
\begin{equation}
	x_{g,l} = \frac{p_{e}e^{\frac{-V_{g,l}}{RT}(p_{e}-p_l^{1atm})}} {H[p_l^{1atm}, T]}.
	\label{eqn-coexist} 
\end{equation}
Eqn.~\ref{eqn-coexist} gives the pressure-mole fraction curve for isobaric equilibrium. This curve, plotted in Fig.~\ref{fig-eqlm-curve}, will be called the isobaric equilibrium curve. 
Fig.~\ref{fig-eqlm-curve} specifically plots the case of oxygen (gas) dissolved in water (liquid), where $H[p_l^{1atm}, T = 25^\textup{o}\textup{C}] = 42,590$~atm and $V_{g,l} = 32$~ml/mol.\cite{Ludw05a}
Eqn.~\ref{eqn-coexist}, whereassumptions for an ideal gas and dilute solution were made, will be less accurate at high mole fractions and high pressures. 
Henry's law is also plotted in Fig.~\ref{fig-eqlm-curve} with $H_{g,l}^c = H[p_l^{1atm}, T = 25^\textup{o}\textup{C}] = 42,590$~atm. 
It is seen that Henry's law is a reasonable approximation to the isobaric equilibrium curve at low mole fractions and pressures. Eqn.~\ref{eqn-coexist} implies that the isobaric equilibrium pressure $p_e$ depends on $T$ and $x_{g,l}$, i.e., $p_e[x_{g,l}, T]$.

\subsection{Non-isobaric equilibrium}
\label{non-isobar}

Equilibrium in a general case where $p_g \neq p_l$ can be written in terms of the isobaric equilibrium pressure $p_{e}[x_{g,l},T]$ as follows
\begin{equation}
	\frac{p_{g}}{p_{e}[x_{g,l},T]} = e^{\frac{V_{g,l}}{RT}(p_l - p_{e}[x_{g,l},T])}.
	\label{eqn-gen-eqlm}
\end{equation}    
Eqn.~\ref{eqn-gen-eqlm} is obtained by using Eqns.~\ref{eqn-eqlm-ideal} and ~\ref{eqn-coexist}. 
To understand Eqn.~\ref{eqn-gen-eqlm} consider the following example. 

\begin{figure}[h]
\centering
\includegraphics[height=6cm]{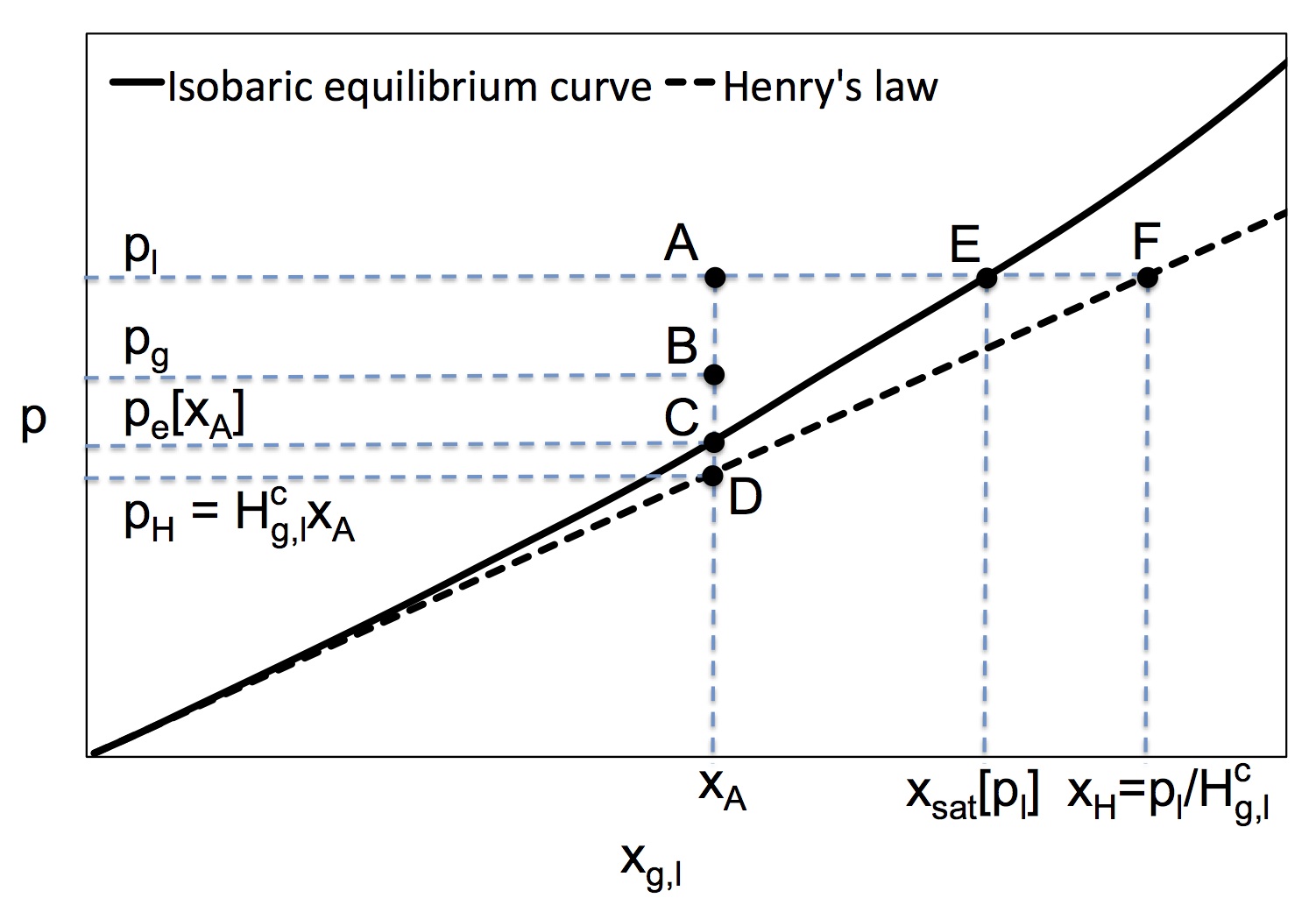}
\caption{A schematic diagram to understand non-isobaric equilibrium between a gas dissolved in a liquid and the same gas adjacent to that liquid.}
\label{fig-analogy}
\end{figure}

Imagine the same configuration as in Fig.~\ref{fig-henry-law} but with a solution pressurized by a mixture of gases instead of only one gas. This configuration is one example of a non-isobaric equilibrium (a second example will be seen in Section 3, where submerged cylindrical pores are considered). The pressure, $p_l$, of the liquid solvent will be the same as the total pressure exerted by all gases on the liquid. Let $x_{g,l} = x_A$ be the mole fraction of one particular gas dissolved in the liquid at pressure $p_l$ -- denoted by point A in Fig.~\ref{fig-analogy}. At equilibrium, what would be the partial pressure $p_g$ of this particular gas in the gaseous mixture above the solution? The answer to this question is provided by Eqn.~\ref{eqn-eqlm-ideal} and denoted by point B in Fig.~\ref{fig-analogy}. Since $p_l \ne p_g$ in this case, the chemical equilibrium between the dissolved gas (point A in Fig.~\ref{fig-analogy}) and the gas phase (point B in Fig.~\ref{fig-analogy}), will be regarded as a non-isobaric equilibrium. 

Given $p_l$ and $x_A$ (point A in Fig.~\ref{fig-analogy}), the gas partial pressure $p_g$ (point B in Fig.~\ref{fig-analogy}) can alternately be obtained by using Eqn.~\ref{eqn-gen-eqlm}. Given $x_A$ (and $T$), the equilibrium pressure $p_e[x_A]$ for an isobaric configuration (Fig.~\ref{fig-henry-law}) is given by Eqn.~\ref{eqn-coexist}. This is denoted by point C in Fig.~\ref{fig-analogy}. Using known values of $p_l$ (point A) and $p_e[x_A]$ (point C), the partial pressure of the gas $p_g$ (point B) can be calculated according to Eqn.~\ref{eqn-gen-eqlm}. 

It is noted that $p_g \ne p_e[x_A]$. If $p_l - p_e$ is not large, then $p_g$ (point B) may be approximated by $p_e[x_A]$ (point C), which in turn may be approximated by $p_H$ at point D by using Henry's law ($p_H = H_{g,l}^c x_A$; Fig.~\ref{fig-analogy}).    

\subsection{Analogy between solubility and phase equilibrium calculations}

Consider Fig.~\ref{fig-analogy} to understand the analogy between solubility and phase equilibrium calculations. The pressure-mole fraction curve for isobaric equilibrium is analogous to the pressure-temperature co-existence (binodal) curve in liquid-vapor phase equilibrium. Eqn.~\ref{eqn-coexist} is analogous to the Clasius-Clapeyron equation (see Eqns.~1-3 in Patankar, 2010\cite{Pata10a} for example). Thus, the isobaric equilibrium pressure $p_e$ in solubility calculations is analogous to the saturation pressure $p_{sat}$ in phase equilibrium calculations.

A gas dissolving into a liquid in case of solubility is analogous to the condensation of a vapor to its liquid in case of phase change. Similarly, the release of a dissolved gas from a liquid (solubility) is analogous to the formation of vapor during boiling (phase change). 

In case of solubility, dissolution of gas is favored above the isobaric equilibrium curve where the liquid is undersaturated, while release of gas from liquid is favored below the isobaric equilibrium curve where the liquid is supersaturated. Analogously, in case of phase change, condensation is favored above the co-existence curve where the liquid is stable, while boiling is favored below the co-existence curve where the liquid is metastable. 

Equating chemical potentials of the gas phase and the dissolved gas (Eqn.~\ref{eqn-gen-eqlm}) in solubility calculations is analogous to equating chemical potentials of the liquid and vapor phases (see, for example, Eqn.~4 in Patankar, 2010\cite{Pata10a}) in phase equilibrium calculations. 

Fig.~\ref{fig-analogy} is a visual aid to organize the calculation process. Specifically, point A with coordinates $(p_l, x_A)$ represents the dissolved gas at mole fraction $x_A$ in a liquid at pressure $p_l$. Point B with coordinates $(p_g, x_A)$ represents the gas phase at pressure $p_g$ that is in chemical equilibrium with the same gas dissolved at mole fraction $x_A$ in the liquid. As long as these interpretations are recognized, Fig.~\ref{fig-analogy}, and the analogy with phase equilibrium, provides a useful framework to conceptualize the calculations. This will be made evident in subsequent sections.





\section{Sustaining gases on rough surfaces}

\subsection{Calculation methodology}

Consider a rough surface with cylindrical pores that is completely submerged in water. Is it possible to sustain a gas in the pores? This question will be analyzed in this section. In this analysis it will be assumed that the length scales are below the capillary length scale of the liquid ($\sim 2.72$~mm for water) so that the effect of gravity is not dominant. 

\begin{figure}[h]
\centering
\includegraphics[height=6cm]{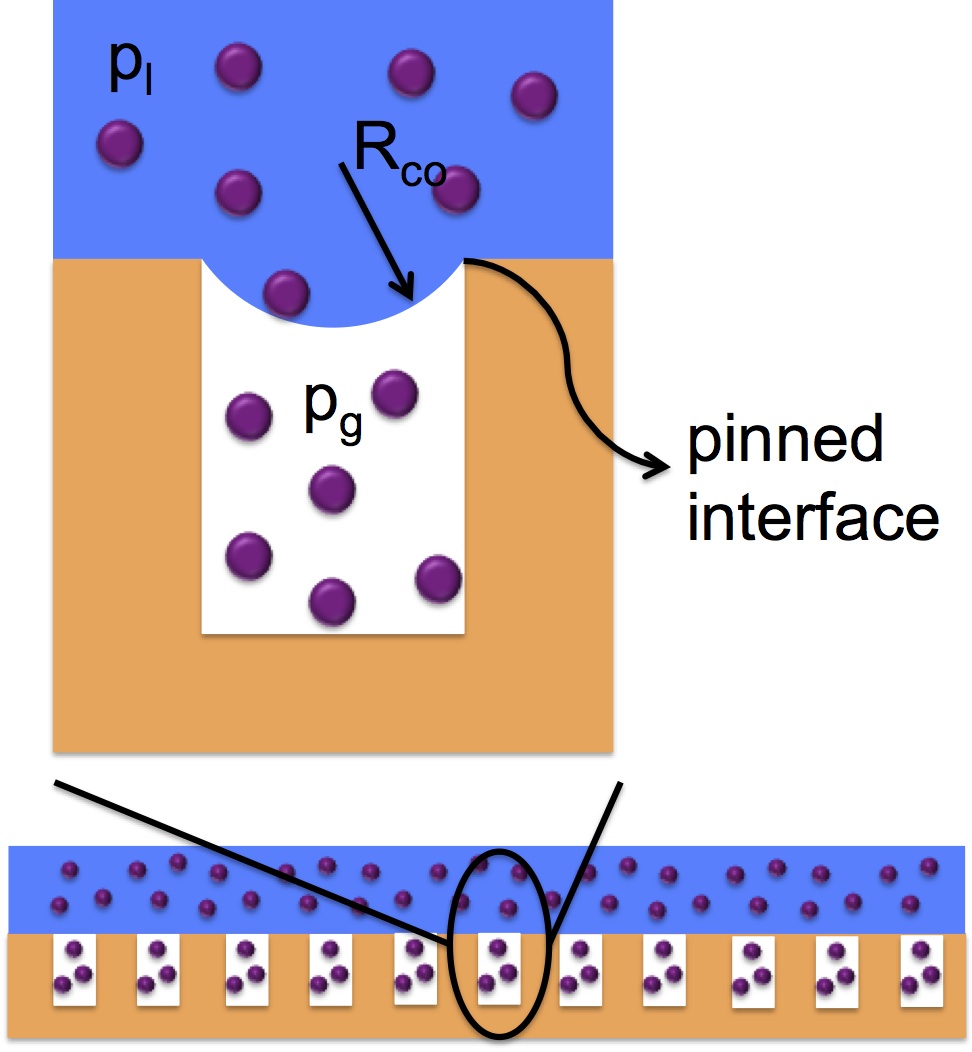}
\caption{Cross-section of a cylindrical pore on a surface submerged in water. Circles denote gas molecules either dissolved in the liquid or inside the pores.}
\label{fig-app}
\end{figure}  

Consider the cross-section of a typical cylindrical pore on a rough surface as shown in Fig.~\ref{fig-app}. Let there be a liquid with a dissolved gas above the pore and the same gas inside the pore. For the gas to be sustained in the pore, there should be thermodynamic equilibrium which constitutes chemical, mechanical, and thermal equilibria. The equilibrium conditions can be derived from energy minimization similar to prior work,\cite{Pata10a} which will not be presented here. The temperature will be assumed uniform everywhere so that thermal equilibrium is ensured. Chemical equilibrium implies that the gas in the pore should have the same chemical potential as the gas dissolved in the liquid above the pore. For simplicity, the liquid will be assumed to be non-volatile. Otherwise chemical equilibrium between liquid and vapor phases of the solvent should be considered.\cite{Pata10a} Mechanical equilibrium implies that the pressure difference between the liquid and the gas in the pore is balanced by the surface tension of the curved interface and that the curved interface remains pinned at the top of the pore (Fig.~\ref{fig-app}).\cite{Pata10a}

Consider Case 1, where the liquid pressure above the pore is $p_l$. Let there be isobaric equilibrium between the gas dissolved in the liquid and the gas inside the pore. For isobaric equilibrium, Eqn.~\ref{eqn-coexist} gives the mole fraction of the dissolved gas in the liquid ($x_{g,l} = x_{sat}[p_l]$) and the gas pressure inside the pore would be equal to the liquid pressure, i.e., $p_g = p_{l}$ (Section~\ref{isobar}). This equilibrium is denoted by point E in Fig.~\ref{fig-analogy}. In this case, the liquid is considered to be saturated with the gas. The dissolved gas and the gas inside the pore would be in chemical equilibrium. Since the liquid and gas pressures are the same, mechanical equilibrium at the interface requires the liquid-gas interface to be flat. If the material of the surface is hydrophobic (i.e. contact angle $> 90^o$) then this flat interface will remain pinned at the corner at the top of the pore\cite{Oliv77a} irrespective of the pore diameter as long as gravitational effect is not dominant. In this case, the gas will be sustained in the pore.

Now consider Case 2, where the liquid pressure is the same as in Case 1, i.e. equal to $p_l$, but the mole fraction of the dissolved gas is $x_{g,l} = x_A < x_{sat}[p_l]$. In this case the liquid is undersaturated with dissolved gas (point A in Fig.~\ref{fig-analogy}). Let the degree of saturation $\phi$ of the dissolved gas denoted by point A in Fig.~\ref{fig-analogy} be quantified by $\phi = x_{g,l}/x_{sat} = x_A/x_{sat}[p_l]$. By imposing chemical equilibrium, the pressure $p_{g}$ of the gas inside the pore can be calculated according to Eqn.~\ref{eqn-eqlm-ideal} or Eqn.~\ref{eqn-gen-eqlm}. This is denoted by point B in Fig.~\ref{fig-analogy}. The dissolved gas (point A) will be in chemical equilibrium with the gas in the pore (point B). This is a case of non-isobaric equilibrium. Mechanical equilibrium requires that the liquid-gas interface will be curved and the radius of curvature $R_{co}$ (Fig.~\ref{fig-app}) of this interface is given by the Young-Laplace equation:
\begin{equation}
	R_{co} = \frac{2\sigma_{lg}}{(p_l - p_g)},
	\label{eqn-young-laplace}
\end{equation}    
where $\sigma_{lg}$ is the liquid-gas interfacial tension. Mechanical equilibrium also requires the liquid-gas interface to remain pinned at the top of the pore in order to sustain gas inside the pore. For this to happen the radius $R$ of the cylindrical pore must satisfy the following pinning condition\cite{Pata10a} 
\begin{align}
	&R < R_{cr},~\textup{where} \nonumber \\
	&R_{cr} = - R_{co} \textup{cos}\theta_e.
	\label{eqn-pinning}
\end{align} 
$\theta_e$ is the equilibrium material contact angle of the surface. Thus, the pore radius should be smaller than the critical radius $R_{cr}$. For Case 1, where the liquid is saturated with the gas, the critical pore radius is infinitely large (under the assumption that gravity is not important). For Case 2, there is a finite value of the critical pore radius. The critical radius will depend on the type of gas dissolved and other conditions such as liquid pressure and the degree of saturation of the dissolved gas, among others. It is, however, possible to simplify the governing equations and obtain estimates for critical pore sizes. This will be discussed in the next section.

It can be verified from the governing equations that as the degree of saturation decreases, i.e., as $\phi$ decreases, the pressure difference, $p_l-p_g$, between the liquid and the gas in the pore will increase for chemical equilibrium to be maintained. This implies smaller radius of curvature $R_{co}$ of the liquid-gas interface. Consequently, the critical radius of the cylindrical pore would be smaller for increasingly undersaturated liquid. 

\subsection{Approximate calculation of the critical pore size}

In this section, approximate calculations for Case 2 of the previous section are presented. The configuration is as in Fig.~\ref{fig-app} and calculations will be done with reference to Fig~\ref{fig-analogy}.

\begin{figure}[h]
\centering
\includegraphics[height=12cm]{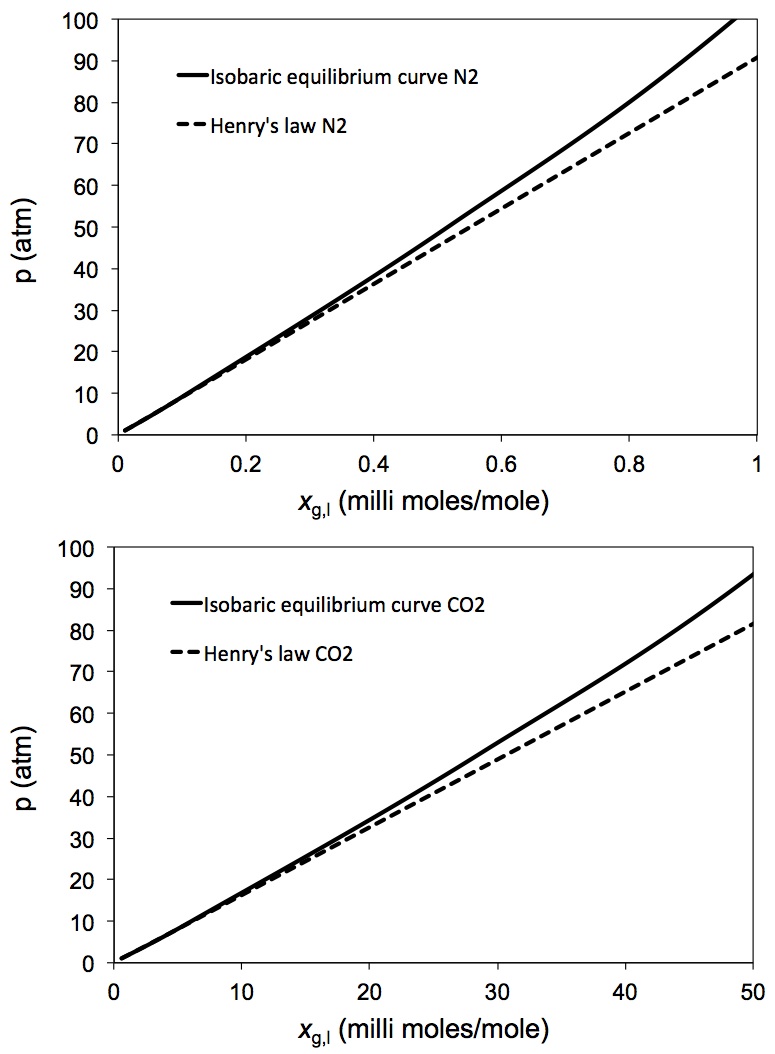}
\caption{Isobaric equilibrium curves (Eqn.~\ref{eqn-coexist}) compared to the equilibrium relation based on Henry's law (Eqn.~\ref{eqn-henry-law}) for nitrogen (top) and carbon dioxide (bottom). $H_{g,l}^c = 90,770$~atm and $V_{g,l} =$ $33.3$~ml/mol for nitrogen,\cite{Ludw05a} and $H_{g,l}^c = 1,630$~atm and $V_{g,l} = $ $37.6$~ml/mol for carbon dioxide.\cite{Gibb71a}}
\label{fig-eqlm-curve-n2-co2}
\end{figure}

Fig.~\ref{fig-eqlm-curve} shows the isobaric equilibrium curve and its comparison with Henry's law for oxygen. Fig.~\ref{fig-eqlm-curve-n2-co2} shows similar curves for nitrogen and carbon dioxide -- two other gases that are also present in air. Carbon dioxide dissolves in much greater quantity in water compared to oxygen and nitrogen. It is seen from Figs.~\ref{fig-eqlm-curve} and~\ref{fig-eqlm-curve-n2-co2}, that for all these gases Henry's law is close to the isobaric curve for pressures as high as 100~atm. Hence, in the following calculations Henry's law will be used to approximate the isobaric equilibrium curve for all gases -- oxygen, nitrogen, and carbon dioxide. In the interest of simplicity, in the following calculations, each of these gases will be considered independently and not as a mixture.   

As noted at the end of Section~\ref{non-isobar}, the gas pressure in the pore, which is denoted by point B in Fig.~\ref{fig-analogy}, will be approximated by point D in Fig.~\ref{fig-analogy}. The error is typically not more than 15\%. For example, in the case of oxygen, it can be verified that the error in calculating oxygen pressure in the pore (difference between $p_g$ at point B and $p_H$ at point D in Fig.~\ref{fig-analogy}) is around 14\% when it is in equilibrium with oxygen that is dissolved in water at 100~atm. (point A in Fig.~\ref{fig-analogy}). While, in the current application the error may be disregarded, this difference cannot be ignored for deep sea organisms and deep diving mammals.\cite{Ludw05a}  

Using approximations above, the following equations are to be noted. First, the degree of saturation $\phi$ of the gas dissolved in the liquid (point A in Fig.~\ref{fig-analogy}) above the pore (Fig.~\ref{fig-app}) can be approximated by 
\begin{equation}
\phi \approx \frac{x_A}{x_{H}},
\label{eqn-undsat}
\end{equation}
where $x_{H}$ is the saturation mole fraction based on Henry's law (point F in Fig.~\ref{fig-analogy}). $x_H$ is used to approximate the saturation mole fraction $x_{sat}[p_l]$ at point E in Fig.~\ref{fig-analogy}. Note that the error between $x_H$ and $x_{sat}[p_l]$ is exaggerated in the schematic in Fig.~\ref{fig-analogy}. 

Second, $x_{H}$ is determined using Henry's law (point F in Fig.~\ref{fig-analogy}) as follows
\begin{equation}
x_{H} = \frac{p_{l}}{H_{g,l}^c}.
\label{eqn-xH}
\end{equation} 

Third, the pressure of the gas in the pore can be approximated by (refer to Fig.~\ref{fig-analogy}) 
\begin{equation}
p_{g} \approx p_e[x_A] \approx p_H = H_{g,l}^c~x_A.
\label{eqn-gas-p}
\end{equation}

Eqns.~\ref{eqn-undsat}-\ref{eqn-gas-p} lead to the following relation
\begin{equation}
\phi \approx \frac{x_A}{x_{H}} \approx \frac{p_{g}}{p_{l}}.
\label{eqn-simple}
\end{equation} 

Use of Eqn.~\ref{eqn-simple} in Eqns.~\ref{eqn-young-laplace} and~\ref{eqn-pinning}, leads to the following equation for the critical pore radius
\begin{equation}
R_{cr} \approx -\frac{2\sigma_{lg}}{p_{l}(1-\phi)} \textup{cos}\theta_e.
\label{eqn-Rcr-approx}
\end{equation} 
The cylindrical pore radius should be less than the critical value obtained from Eqn.~\ref{eqn-Rcr-approx} to sustain gas within a cylindrical pore. Note that, with the approximations used, the critical radius does not depend on the type of gas being considered once the degree of saturation $\phi$ is known. 

\begin{figure}[h]
\centering
\includegraphics[height=6cm]{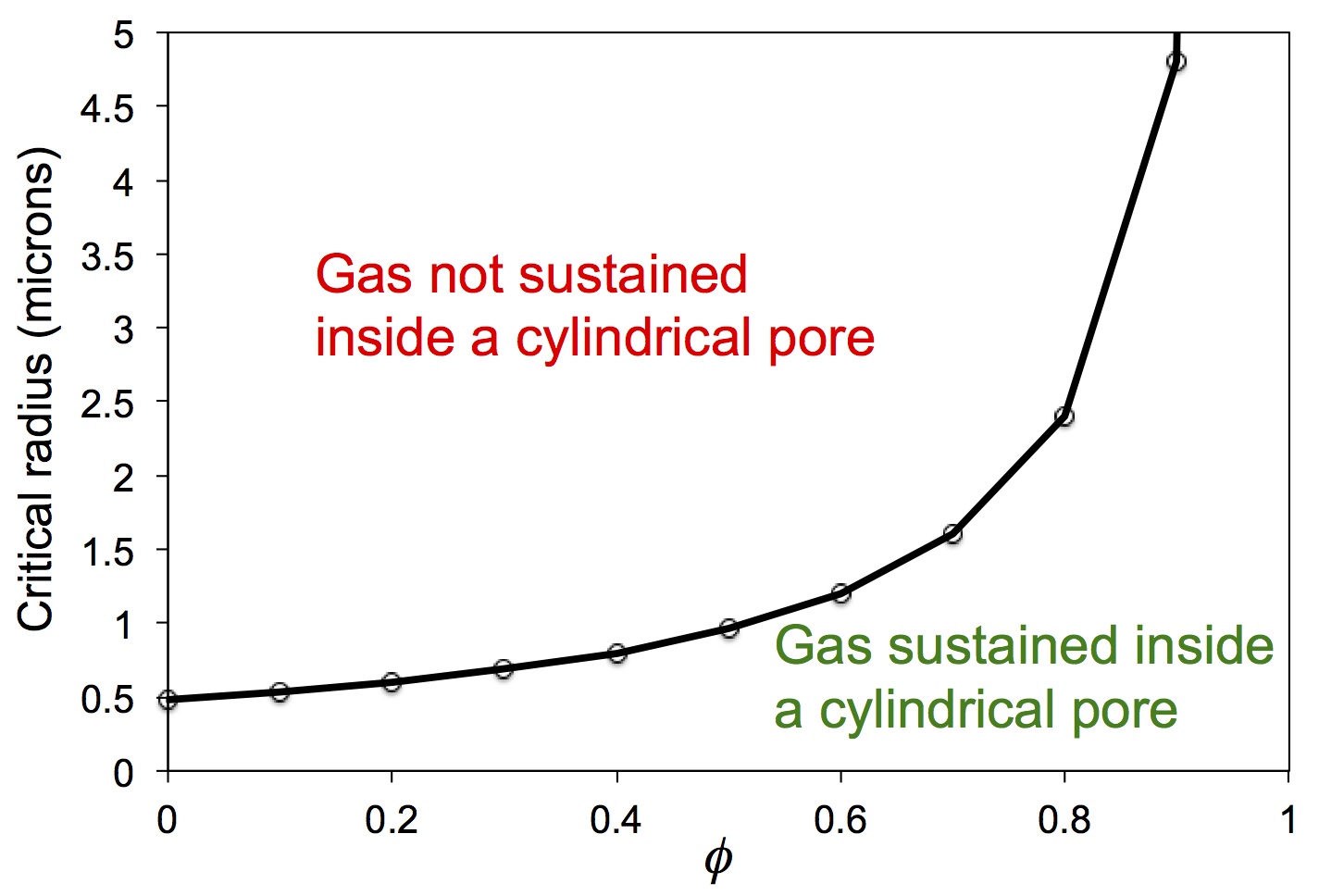}
\caption{Critical pore radius to sustain gas in cylindrical pores on a surface that is submerged in water. Pore radii should be below the critical radius to sustain gas. $\phi$ is the degree of saturation of the gas dissolved in liquid water. Liquid water is assumed to be at 1~atm., $\sigma_{lg} = 72$~mN/m, and $\theta = 110^o$.}
\label{fig-rcr}
\end{figure}

Fig.~\ref{fig-rcr} shows the critical pore radius for different degrees of saturation of gases dissolved in water at $p_{l} = 1$~atm. The surface tension of the water-gas interface is $\sigma_{lg} = 72$~mN/m at room temperature and $\theta = 110^o$ is considered (typical for hydrophobic chemical coatings on rough surfaces). 
Fig.~\ref{fig-rcr} shows that the smallest critical pore radius at 1~atm. water pressure is 480~nm corresponding to $\phi = 0$, i.e. no dissolved gases in water. In other words pore diameters of less than a micron are necessary to prevent the liquid from wetting the pores irrespective of the degree of saturation of dissolved gases. Similar analysis can be extended to pillar or other geometries without fundamental difficulty.\cite{Pata10b} Calculations predict that pillar spacing of less than a micron would be required to prevent wetting of rough surfaces submerged in water. This is consistent with experimental results where it is observed that surfaces remain dry under water when roughness length scales are below micron scale, whereas surfaces with tens of micron scale roughness typically get wet in 3-4 days for liquid pressures around 1~atm.\cite{Gova09a, Leec11a, Poet10a, Sama11a, Balm11a}

It also follows from Eqn.~\ref{fig-rcr} that the critical radius decreases with increasing liquid pressure for the same degree of saturation. For example, the critical pore radius would reduce from 480~nm to 48~nm if the water pressure were increased from 1~atm. to 10~atm.






\section{Conclusions}

Thermodynamics of sustaining gases (e.g. air) on rough surfaces submerged in water is considered. Governing equations and a solution methodology to determine critical condition to sustain gases is presented. It is shown that roughness spacings of less than micron are essential to sustain gases on surfaces submerged in water at atmospheric pressure. This is consistent with prior empirical data.

In the analysis presented here, the liquid, into which the rough surfaces would be submerged, was considered to be either saturated or undersaturated with dissolved gases. Extension to supersaturated cases can be done without difficulty. In that case the liquid pressure would be lower than the gas pressure at equilibrium. Consequently, sustaining gases would be possible even with hydrophilic surfaces. However, if the liquid is undersaturated with gas then hydrophilic surfaces submerged in it would not be able to sustain gases inside cylindrical pores -- reentrant roughness geometry would be necessary.\cite{Herm00a, Tute07a} In general superhydrophilic surfaces would promote wetting of submerged surfaces.   

\section*{Acknowledgements} Support from the Institute for Sustainability and Energy at Northwestern (ISEN) is gratefully acknowledged. N.A.P. thanks Paul Jones for helpful suggestions. 




\bibliography{./references} 

\begin{thebibliography}{10}
\expandafter\ifx\csname urlstyle\endcsname\relax
  \providecommand{\doi}[1]{doi:\discretionary{}{}{}#1}\else
  \providecommand{\doi}{doi:\discretionary{}{}{}\begingroup
  \urlstyle{rm}\Url}\fi

\bibitem{Gova09a}
Govardhan, R.~N., Srinivas, G.~S., Asthana, A. \& Bobji, M.~S., 2009 Time
  dependence of effective slip on textured hydrophobic surfaces.
\newblock \emph{Phys. Fluids} \textbf{21}, 052001.

\bibitem{Leec11a}
Lee, C. \& Kim, C.~J., 2011 Underwater restoration and retention of gases on
  superhydrophobic surfaces for drag reduction.
\newblock \emph{Phys. Rev. Lett.} \textbf{106}, 014502.

\bibitem{Mcha09a}
McHale, G., Shirtcliffe, N.~J., Evans, C.~R. \& Newton, M.~I., 2009 Terminal
  velocity and drag reduction measurements on superhydrophobic spheres.
\newblock \emph{Appl. Phys. Lett.} \textbf{94}, 064104.

\bibitem{Dhir06a}
Dhir, V.~K., 2006 Mechanistic prediction of nucleate boiling heat transfer -
  achievable or a hopeless task?
\newblock \emph{Journal of Heat Transfer-Transactions of the ASME}
  \textbf{128}, 1--12.

\bibitem{Poet10a}
Poetes, R., Holtzmann, K., Franze, K. \& Steiner, U., 2010 Metastable
  underwater superhydrophobicity.
\newblock \emph{Phys. Rev. Lett.} \textbf{105}, 166104.

\bibitem{Sama11a}
Samaha, M.~A., Ochanda, F.~O., Tafreshi, H.~V., Tepper, G.~C. \& Gad-el Hak,
  M., 2011 In situ, noninvasive characterization of superhydrophobic coatings.
\newblock \emph{Rev. Sci. Instrum.} \textbf{82}, 045109.

\bibitem{Balm11a}
Balmert, A., Bohn, H.~F., Ditsche-Kuru, P. \& Barthlott, W., 2011 Dynamic air
  layer on textured superhydrophobic surfaces.
\newblock \emph{J. Morphol.} \textbf{272}, 442--451.

\bibitem{Marm06a}
Marmur, A., 2006 Underwater superhydrophobicity: Theoretical feasibility.
\newblock \emph{Langmuir} \textbf{22}, 1400--1402.

\bibitem{Flyn08a}
Flynn, M.~R. \& Bush, J. W.~M., 2008 Underwater breathing: The mechanics of
  plastron respiration.
\newblock \emph{J. Fluid Mech.} \textbf{608}, 275--296.

\bibitem{Vaka13a}
Vakarelski, I.~U., Chan, D.~Y., Marston, J.~O. \& Thoroddsen, S.~T., 2013
  Dynamic air layer on textured superhydrophobic surfaces.
\newblock \emph{Langmuir} \textbf{29}, 11074--11081.

\bibitem{Pata10a}
Patankar, N.~A., 2010 Supernucleating surfaces for nucleate boiling and
  dropwise condensation heat transfer.
\newblock \emph{Soft Matter} \textbf{6}, 1613--1620.

\bibitem{Pata10b}
Patankar, N.~A., 2010 Vapor stabilizing substrates for superhydrophobicity and
  superslip.
\newblock \emph{Langmuir} \textbf{26}, 8783--8786.

\bibitem{Vaka12a}
Vakarelski, I.~U., Patankar, N.~A., Marston, J.~O., Chan, D. Y.~C. \&
  Thoroddsen, S.~T., 2012 Stabilization of leidenfrost vapour layer by textured
  superhydrophobic surfaces.
\newblock \emph{Comp. Biochem. Phys. A} \textbf{489}, 274--277.

\bibitem{Jone14a}
Jones, P.~R., Hao, X., Cruz-Chu, E.~R., Rykaczewski, K., Nandy, K., Schutzius,
  T.~M., Varanasi, K.~K., Megaridis, C.~M., Walther, J.~H., Koumoutsakos, P.
  \emph{et~al.}, 2014 Sustaining dry surfaces under water.
\newblock \emph{arXiv:1409.8218} .

\bibitem{Mode83a}
Modell, M. \& Reid, R.~C., 1983 \emph{Thermodynamics and Its Applications}.
\newblock Englewoods Cliffs, NJ.: Prentice-Hall.

\bibitem{Ludw05a}
Ludwig, H. \& MacDonald, A.~G., 2005 The significance of the activity of
  dissolved oxygen, and other gases, enhanced by high hydrostatic pressure.
\newblock \emph{Comp. Biochem. Phys. A} \textbf{140}, 387--395.

\bibitem{Oliv77a}
Oliver, J.~F., Huh, C. \& Mason, S.~G., 1977 Resistance to spreading of liquids
  by sharp edges.
\newblock \emph{Journal of Colloid and Interface Science} \textbf{59},
  568--581.

\bibitem{Gibb71a}
Gibbs, R.~E. \& Van~Ness, H.~C., 1971 Solubility of gases in liquids in
  relation to the partial molar volumes of the solute. carbon dioxide-water.
\newblock \emph{Ind. Eng. Chem. Fundam.} \textbf{10}, 312--315.

\bibitem{Herm00a}
Herminghaus, S., 2000 Roughness-induced non-wetting.
\newblock \emph{Europhys. Lett.} \textbf{52}, 165--170.

\bibitem{Tute07a}
Tuteja, A., Choi, W., Ma, M., Mabry, J.~M., Mazzella, S.~A., Rutledge, G.~C.,
  McKinley, G.~H. \& Cohen, R.~E., 2007 Designing superoleophobic surfaces.
\newblock \emph{Science} \textbf{318}, 1618--1622.

\end{thebibliography}
\bibliographystyle{prsb}

\end{document}